\documentclass[12pt]{article}

\usepackage{epic}

\usepackage{latexsym}
\usepackage{amsmath}
\usepackage{epsfig}
\usepackage{amssymb}
\usepackage{enumerate}
\usepackage{centernot}
\usepackage{nicefrac}
\usepackage[mathscr]{euscript}
\usepackage{geometry} 
\newtheorem{theorem}{Theorem}
\newtheorem{lemma}[theorem]{Lemma} 

\newtheorem{proposition}[theorem]{Proposition}

\newcommand{\PP}{{\mathbb P}}
\newcommand{\EE}{{\mathbb E}}

\begin{document}
\title{Similarities as Evidence for Common Ancestry--A Likelihood Epistemology}
\date{\today} 
\maketitle

\begin{center}
\noindent Elliott Sober\footnote{ Philosophy Department, University of Wisconsin, Madison, Wisconsin, USA.} and Mike Steel\footnote{Biomathematics Research Centre, University of Canterbury, Christchurch, New Zealand.}
\end{center}

\thanks{}

\tableofcontents

\newpage

\begin{abstract}
\noindent Darwin claims in the {\em Origin} that similarity is evidence for common ancestry, but that  adaptive similarities are ``almost valueless'' as evidence.  This claim seems reasonable for some adaptive similarities but not for others. 
  Here we clarify and evaluate these and related matters by using the law of likelihood as an analytic tool and by considering mathematical models of three evolutionary processes  -- directional selection, stabilizing selection, and drift.  Our results apply both to Darwin's theory of evolution and to modern evolutionary biology.     
\end{abstract}



\section{Introduction}

In the last paragraph of the {\em Origin}, Darwin (1859, p. 490)
says that, in the beginning, life was breathed ``into a few forms, or into one.'' The caution embodied in ``one or a few'' is not to be found in present-day biology, which embraces the idea of {\em universal} common ancestry.  Darwin tentatively reaches towards that stronger thesis a few pages earlier:\\

\begin{quote}
... I believe that animals have descended from at most only four or five progenitors, and plants from an equal or lesser number.  Analogy would lead me one step further, namely to the belief that all animals and plants have descended from some one prototype.  But analogy may be a deceitful guide.  Nevertheless all living things have much in common, in their chemical composition, their germinal vesicles, their cellular structure, and their laws of growth and reproduction.  We see this even in so trifling a circumstance as that the same poison often similarly affects plants and animals; or that the poison secreted by the gall-fly produces monstrous growths on the wild rose or oak-tree.  Therefore I should infer from analogy that probably all organic beings which have ever lived on this 	earth have descended from some one primordial form, into which life was first breathed. (Darwin 1859, p. 484)
\end{quote}

\vspace{0.5cm} 

\noindent Darwin's idea that universal common ancestry is supported by the fact that ``all living things have much in common'' is an instance of a broader principle: when two or more taxa have trait $X$, this similarity favors the hypothesis of common ancestry over the hypothesis of separate ancestry.  

Darwin advances a second epistemological thesis about common ancestry -- that some similarities provide stronger evidence for common ancestry than others: \\

\begin{quote}
... adaptive characters, although of the utmost importance to the welfare of the being, are almost valueless to the systematist.  For animals belonging to two most distinct lines of descent, may readily become adapted to similar conditions, and thus assume a close external resemblance; but such resemblances will not reveal -- will rather tend to conceal their blood-relationship to their proper lines of descent. (Darwin 1859, p. 427)
\end{quote}

\vspace{0.5cm} 

\noindent On the next page, he gives the example of the ``shape of the body and the fin-like limbs'' found in whales and fishes; these are ``adaptations in both classes for swimming through water'' and thus provide almost no evidence that the two groups have a common ancestor.  

Although Darwin's principle -- that adaptive similarities provide scant evidence for common ancestry -- sounds right when it is applied to this example, there are other examples in which it sounds wrong.  Darwin describes one of them:\\

\begin{quote}
The framework of bones being the same in the hand of a man, wing of a bat, fin of the porpoise, and leg of the horse -- the same number of vertebrae forming the neck of the giraffe and of the elephant, -- and innumerable other such facts, at once explain themselves on the theory of descent with slow and slight successive modifications. The similarity of pattern in the wing and leg of a bat, though used for such different purposes, -- in the jaws and legs of a crab, -- in the petals, stamens, and pistils of a flower, is likewise intelligible on the view of the gradual modification of parts or organs, which were alike in the early progenitor of each class. (Darwin 1859, p. 479)
\end{quote}

\vspace{0.5cm} 

\noindent The shared ``framework of bones'' seems to be evidence for common ancestry, and yet  this morphology seems to be useful in the different groups (Lewens 2015).  So which epistemological principle is right -- that {\em all} adaptive similarities provide only meager evidence for common ancestry, or that {\em some} adaptive similarities provide weak evidence while others provide strong?  If the latter, how can the one sort of adaptive similarity be separated from the other?

	Darwin's prose suggests an answer to this last question: perhaps the shared framework of bones is strong evidence for common ancestry because it is used for different purposes in these different groups.  This suggestion seems to separate the torpedo-shape of whales and fish from the limb morphology of human beings, bats, porpoises, and horses.  
However, there is a more modern example that	should give us pause about this proposal.   Crick (1957)
argued that the universality of the genetic code is strong evidence for common ancestry. Modern biology has retained his conclusion even though we now know that the genetic code isn't universal; it is {\em nearly} universal, with almost all groups of organisms using one code and a few others using codes that are very similar, but not identical, to the one (Knight, Freeland, and Landweber 2001).
The prevalent genetic code provides strong evidence for common ancestry even though it has the same purpose in all the living things that have it.

\section{The likelihood framework}
To sort out Darwin's ideas concerning evidence for common ancestry, we need two concepts -- one qualitative, the other quantitative.  The former is provided by the law of likelihood (Hacking 1965):

\bigskip

\noindent {\bf (Qual) } Observation $O$ favors hypothesis $H_1$ over hypothesis $H_2$ if and only if 
$$Pr(O|H_1) > Pr(O|H_2).$$

\noindent We will use this epistemological principle when we describe a fairly general set of assumptions in Section 3 that entails that

\begin{equation*}
\begin{aligned}
&Pr(\mbox{taxa $A$ and $B$ have trait $X | A$ and $B$ have a common ancestor) $>$} \\
&Pr(\mbox{taxa $A$ and $B$ have trait $X | A$ and $B$ do not have a common ancestor)}. 
\end{aligned}
\end{equation*}

\vspace{0.5cm}

\noindent Qual takes this inequality to mean that the similarity connecting $A$ and $B$ favors the hypothesis of common ancestry (CA) over the hypothesis of separate ancestry (SA).  However, Qual does not provide the resources for describing how the type of evolutionary process affects the degree to which the similarity favors CA over SA.  For this purpose we will use\\

\noindent  {\bf (Quant)}
The degree to which $O$ favors $H_1$ over $H_2$ is given by the likelihood ratio $$\frac{Pr(O|H_1)}{Pr(O|H_2)}.$$\\

\noindent Given that taxa $A$ and $B$ share trait $X$, we will compare how strongly this similarity favors CA over SA when $M$ is the evolutionary process that governed the evolution of $X$ with how strongly the similarity favors CA over SA when $N$ is the process at work. This will involve comparing two likelihood ratios:

$$\frac{Pr_M (\mbox{$A$ and $B$ have trait $X |  CA$})}{Pr_M (\mbox{$A$ and $B$ have trait $X |  SA$})}  > 
  \frac{Pr_N (\mbox{$A$ and $B$ have trait $X |  CA$})}{Pr_N (\mbox{$A$ and $B$ have trait $X |  SA$})}.$$

\bigskip

\noindent We consider which pairs of processes are related by this inequality in Sections 4-6.

\section{A sufficient condition for a similarity to favor common ancestry over separate ancestry}

Inspired by Reichenbach's (1956)
discussion of his principle of the common cause, we here describe a sufficient condition for a dichotomous trait shared by taxa $A$ and $B$ to favor the hypothesis that $A$ and $B$ have a common ancestor over the hypothesis that they do not.  Figure \ref{fig1_sober} depicts the two hypotheses.  

\bigskip

\begin{figure}[htb]
\centering
\includegraphics[scale=1.2]{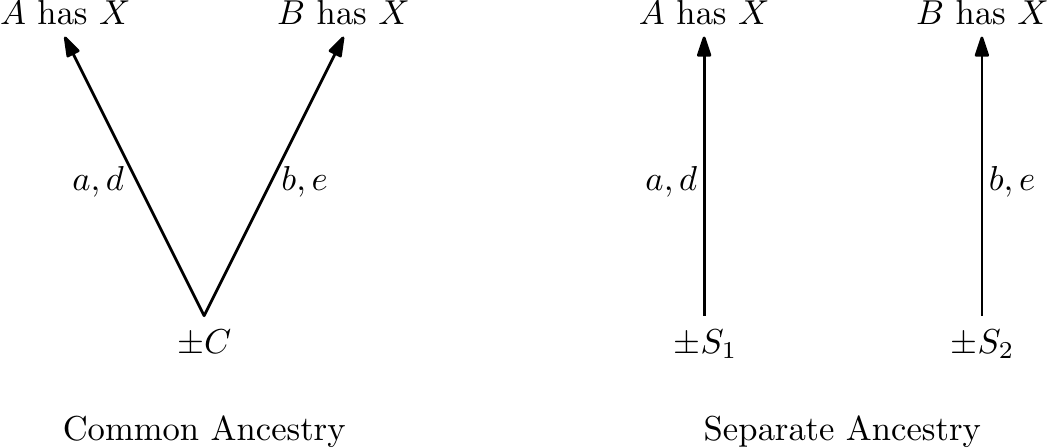}
\caption{Taxa $A$ and $B$ are observed to have trait $X$.  Does this observation favor Common Ancestry over Separate Ancestry?  The probabilistic parameters $a, d, b,$ and $e$ are explained in the text.}
\label{fig1_sober}
\end{figure}

Notice that the states of $A$ and $B$ are described in the figure ($A$ and $B$ are both observed to have trait $X$), but the states of the ancestors postulated by the two hypotheses are not.  The ancestors are represented by variables that take one of two values; the ``+'' value means that the postulated ancestor had trait $X$ while the ``$-$'' value means that the ancestor lacked $X$.   Next to the branches in Figure 1 are lower-case letters that denote probabilities.  These are

\begin{equation*}
\begin{aligned}
	a &= Pr(A \mbox{ has } X | C \mbox{ has } X) \\
  d &= Pr(A \mbox{ has } X | C \mbox{ lacks } X) \\
  b &= Pr(B \mbox{ has } X | C \mbox{ has } X) \\
  e &= Pr(B \mbox{ has } X | C \mbox{ lacks } X) 
\end{aligned}
\begin{aligned}
	&= Pr(A \mbox{ has } X | S_1 \mbox{ has } X) 	\\
	&= Pr(A \mbox{ has } X | S_1 \mbox{ lacks } X) 	 \\
	&= Pr(B \mbox{ has } X | S_2 \mbox{ has } X) 	\\
	&= Pr(B \mbox{ has } X | S_2 \mbox{ lacks } X) 
\end{aligned}
\end{equation*}

\noindent There is one more probability that we need, but it isn't in the figure: 
	
$$c = Pr(C \mbox{ has } X) = Pr(S_1 \mbox{ has } X) = Pr(S_2 \mbox{ has } X)$$

\bigskip

\noindent The fact that the parameters that attach to the Common Ancestry hypothesis also attach to the Separate Ancestry hypothesis represents an assumption:\\

\begin{quote}
\noindent{\em Assumption 1} (cross-model homogeneity):
The probability that a taxon has a trait does not depend on whether the Common Ancestry or the Separate Ancestry hypothesis is true. \\ 
\end{quote}

\noindent We add four more assumptions:  

\begin{quote}
\noindent{\em Assumption 2 }(screening-off): 
\begin{enumerate}[(i).]
\item  $Pr(A \mbox{ and } B \mbox{ have } X | \pm C) = Pr(A \mbox{ has } X | \pm C)Pr(B \mbox{ has } X | \pm~C).$
\item  $Pr(A \mbox{ and  } B \mbox{ have }  X | \pm S_1 \& \pm S_2)  = Pr(A \mbox{ has } X | \pm S_1 \& \pm~S_2)Pr(B \mbox{ has } X | \pm~S_1\&\pm~S_2).$
\item $Pr(A \mbox{ has } X | \pm S_1 \& \pm S_2) =Pr(A \mbox{ has } X | \pm S_1)$ and 

$Pr(B \mbox{ has } X | \pm~S_1\& \pm S_2) = Pr(B \mbox{ has } X | \pm S_2)$
\end{enumerate}

\noindent{\em Assumption 3} (non-extreme probabilities): $0<a,b,c,d,e<1$.

\noindent{\em Assumption 4} (ancestor independence):  $Pr(\pm S_1 \& \pm S_2) = Pr(\pm S_1)\Pr(\pm S_2).$

\noindent{\em Assumption 5} (cross-branch homogeneity):
$(a-d)$ and $(b-e)$ are either both positive or both negative.  The common ancestor's having trait $X$ must make a difference (either positive or negative) in the probability that one of its descendants will have trait $X$, and it must make a difference of the same sign in the probability that the other descendant will have trait $X$.\\
\end{quote}

\noindent Notice that Assumption 5 is qualitative, not quantitative; it does not say that $a=b$ and $d=e$.  
\begin{quote}
	Simple algebra suffices to establish the following:

If taxa $A$ and $B$ have trait $X$ (where there are just two trait values) and assumptions 1-5 are true, then 
$$Pr(\mbox{taxa } A \mbox{ and } B \mbox{ have trait } X | CA) > Pr(\mbox{taxa } A \mbox{ and } B \mbox{ have trait } X | SA).$$
\end{quote}

\noindent It does not matter whether the evolutionary process at work in a branch is selection or drift or whether the same process is at work in different branches.  The five assumptions are not {\em a priori} true, but they are very general; all the mathematical models for the evolution of a dichotomous trait that biologists now use  in phylogenetic inference obey these assumptions (Lemey, Salemi and Vandamme 2009).
We note, in particular, that Assumption 5 holds for any Markov model of the evolution of a dichotomous trait since such models obey a ``backwards inequality'': $ Pr_t(\mbox{descendant has } X | \mbox{ ancestor has } X) > Pr_t(\mbox{descendant has } X | \mbox{ ancestor lacks } X)$, for any finite amount of time $t$ between ancestor and descendant (Sober 2008).  This inequality applies to each branch and so the cross-branch homogeneity assumption is satisfied.

What happens if the evolving trait has $n$ values $X_1, X_2, \cdots, X_n$? Assumptions 1--4 modify in a straightforward way, by simply replacing each of the two states `has $X$' (denoted by the $+$) and `lacks $X$' (denoted by $-$) by each of the $n$ possible states (so, for example, the modification of Assumption 2(ii) will represent $n^2$ statements rather than just four when $n>2$). 
However, the application of Assumption 5 merits spelling out.  What is needed is this: 

\begin{quote}
There exist states $X_i, X_j$ (distinct) and $X_k$ (possibly equal to $X_i$ or $X_j$) so that 
changing the state of the common ancestor from $X_i$ to $X_j$  raises
the probability that one descendant will be in state $X_k$. And for any  two distinct states $X_l$ and $X_m$,  if changing the state of the common ancestor from $X_l$ to $X_m$ raises the probability that one descendant will be in state $X_k$, the change also will raise the probability that the other descendant will be in state $X_k$.  
 \end{quote}
In this case we have the following result, a brief proof of which  is provided in the Appendix.

\begin{proposition}
\label{reich}  
Under assumptions A1--A5, extended to allow $n$ states,
$$Pr(\mbox{taxa $A$ and $B$ are in state $X_k | CA$}) > Pr(\mbox{taxa $A$ and $B$ are in state $X_k | SA$}).$$
\end{proposition}

Although the argument for Proposition~\ref{reich} goes through, the fact that the evolving trait isn't dichotomous opens the door to possible violations of Assumption 5.  An example is depicted in the accompanying table.  In the branch leading to taxon $A$, there is strong selection for $X_2$ if the common ancestor $C$ is in state $X_1$, but strong stablizing selection prevents $X_2$ from evolving if $C$ is in state $X_3$ and also prevents $X_3$ from evolving if $C$ is in state $X_2$.   In the lineage leading to taxon $B$, the situation is just the reverse: there is strong selection for $X_2$ if the ancestor is in state $X_3$, but stabilizing selection prevents $X_1$ from evolving into $X_2$, and also prevents $X_2$ from evolving into $X_1$.  This difference in the processes governing trait evolution in the two branches gives rise to the two probabilistic inequalities described in the table; together, they violate Assumption 5.  Suppose that ancestors (in both the common ancestry and the separate ancestry models) have $X_1$, $X_2$, and $X_3$ with probabilities 0.49, 0.02, and 0.49, respectively.   The result is that if taxa $A$ and $B$ are both in state $X_2$, this similarity will favor separate ancestry over common ancestry.  The likelihood of the common ancestry hypothesis is approximately 0.02, whereas the likelihood of the separate ancestry hypothesis is about $(0.51)^2$.

\begin{figure}[htb]
\centering
\includegraphics[scale=0.85]{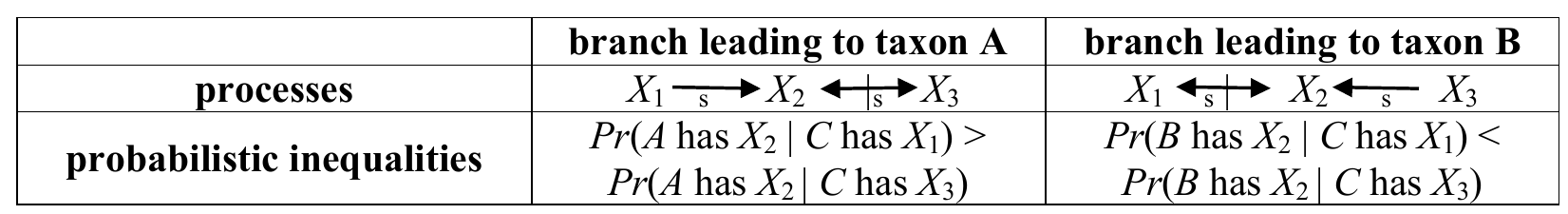}
\label{fig1_sober}
\end{figure}

\vspace{0.5cm}
  
Our analysis in this section concerns two taxa.  If there are more than two, the taxa can differ in how closely related they are to each other under the common ancestry hypothesis.  We address this complication in Section 7.  

Our results so far lend support to Darwin's intuition that similarity is evidence for common ancestry.   The assumptions needed to derive this result aren't {\em a priori} but they are very widely satisfied.\footnote{For discussion of the relation of this argument to Reichenbach's principle of the common cause, and for examples outside of evolutionary biology in which similarity can be evidence favoring a separate cause model over a common cause model, see Sober (2015).}

\section{The $\nicefrac{1}{p}$ criterion and its limitations}
\label{1psec}

We now turn to the question of which processes strengthen the evidence that a similarity provides for common ancestry and which processes weaken that evidence.  We begin with a simple argument presented by Sober and Steel (2014a). The likelihood ratio of CA to SA can be expanded as follows:

$$\frac{Pr_{CA}(A \mbox{ and } B \mbox{ have trait } X)}{Pr_{SA}(A \mbox{ and } B \mbox{ have trait } X)}  = 
\frac{Pr_{CA}(A \mbox{ has trait } X | B \mbox{ has trait } X)Pr_{CA}(B \mbox{ has trait } X)}{Pr_{SA}(A \mbox{ has trait } X) Pr_{SA}(B \mbox{ has trait } X)} .$$

\vspace{0.5cm}

\noindent If we use Assumption 1 above, that

	$$Pr_{CA}(B \mbox{ has trait } X) = Pr_{SA}(B \mbox{ has trait } X) = p$$
	
	\bigskip

\noindent and assume further that the evolutionary process is uniform (meaning that simultaneous branches have the same probabilities of changing state), so that 

$$Pr_{CA}(A \mbox{ has trait } X) = Pr_{CA}(B \mbox{ has trait }  X),$$

\bigskip

\noindent the likelihood ratio becomes:
\begin{equation}
\label{condeq}
\frac{Pr_{CA}(A \mbox{ and } B \mbox{ have trait } X)}{Pr_{SA}(A \mbox{ and } B \mbox{ have trait } X)} = \frac{Pr_{CA}(A \mbox{ has trait } X |  B \mbox{ has trait } X)}{p}
\end{equation}
Suppose, finally, that if $A$ and $B$ have a common ancestor, then the amount of time between $A$ and $B$ and their most recent common ancestor is very small.  This entails that the likelihood ratio is approximately $\nicefrac{1}{p}$.

	Given that this likelihood ratio gets bigger as $p$ gets smaller, there is a simple argument for an implication of Darwin's thesis that adaptive similarities provide little evidence for common ancestry.  The argument does not describe the absolute amount of evidence that adaptive similarities provide, but it does say the following:  if the value for $p$ when $X$ is adaptive is greater than the value for $p$ when $X$ is neutral or deleterious, then neutral and deleterious similarities provide stronger evidence for common ancestry than adaptive similarities do.  We will see in what follows that there are counterexamples to the consequent of the conditional just stated; there are adaptive similarities that provide stronger evidence for common ancestry than neutral similarities provide.  Even so, the $\nicefrac{1}{p}$ argument is a good starting point.\footnote{Imperfect approximations of the $1/p$ argument are presented in Sober (2008, pp. 297--305; 2011, p. 30).}


The argument has two limitations.   The first, that the argument is formulated for just two taxa, will be removed in Section~\ref{beyond}.   The second limitation is the assumption that if $A$ and $B$ have a common ancestor, they have a very recent common ancestor.  This limitation can be lifted by considering a Markovian process of character state evolution.  For computational convenience, in this paper we consider the simplest model that allows different traits to have different probabilities, namely the 
 {\em equal input model}.  This model, for an $n$-state character $X$, says that for all states $j,k$ different from $i$, $Pr(\mbox{descendant has }X_i | \mbox{ancestor has }X_j) = Pr(\mbox{descendant has } X_i | \mbox{ancestor has }X_k)$ (Semple and Steel 2003).\footnote{For 4-state characters this is sometimes called `Felsenstein's 1981' model or the `Tajima-Nei' equal input model.}   Except where we consider directional selection, we will suppose the equal input model is stationary.  Any stationary equal input model involving any number of states entails the following succinct representation of the likelihood ratio (Sober and Steel, 2014a):

\begin{equation}
\label{sobeq}
LR^{p}_{CA/SA} = 1+ \frac{(1-p)}{p}e^{-2rt}.
\end{equation}
Here the most recent common ancestor of $A$ and $B$ postulated by the CA model is $t$ units of time in the past, $p$ is the stationary probability of trait $X$, and $r$ is a scaled rate of substitution between states.  Notice that the likelihood ratio in (\ref{sobeq}) is greater than unity when $t$ is any finite positive number (though it asymptotically approaches unity as $t$ is made large) and that the ratio is made large by making $p$ small. Later, we will see that Eqn. (\ref{sobeq}) also falls out as a corollary of Proposition~\ref{thm1}.

\section{Directional selection versus drift}
\label{dirsec}
  Under CA, suppose that the root of the 2-taxon tree is in state $X$ with probability $q$, and that $s$ is the stationary probability for state $X$. Thus the probability $p$ that a present day taxon is in state $X$  lies between $q$ and $s$, so  either $p=q=s$ (neutrality) or $q<p<s$ (selection for trait $X$) or $s<p<q$ (selection against trait $X$) when $t>0$.
In the equal input model, the probability of being in state $X$ if the process was in state $Y$ at $t$ units in the past is 
$s +(1-s)e^{-rt}$ when $Y=X$ and $s(1-e^{-rt})$ for $Y \neq X$.  Therefore:
$$p=q[s + (1-s)e^{-rt}] + (1-q)[(1-e^{-rt}],$$
which simplifies  to the relationship:
\begin{equation}
\label{pq}
p = s(1-e^{-rt}) + qe^{-rt}.
\end{equation}
For $t$ small, $p$ is close to $q$ (with equality at $t=0$) and for $t$ large, $p$ is close to $s$ (with equality in the limit as $t \rightarrow \infty$). 
It is convenient to think of $s$ and $q$ as given, with $p$  determined by these quantities (and $r,t$) via Eqn. (\ref{pq}). Consider now the likelihood ratio of CA to SA under directional selection, which we denote by $LR^{DS}_{CA/SA}$. Applying Eqn. (\ref{pq}) in the denominator, we have:
\begin{equation}
\label{dirbig}
LR^{DS}_{CA/SA} = \frac{q[s+ (1-s)e^{-rt}]^2 + (1-q)[s(1-e^{-rt})]^2}{[s(1-e^{-rt}) + qe^{-rt}]^2}.
\end{equation}
If any two of $q, s, p$ are equal then all three are, in which case directional selection disappears, and from (\ref{sobeq}) 
we obtain the likelihood ratio  of common ancestry to separate ancestry when there is drift (D), which we denote by $LR^{D}_{CA/SA}$. That is:
 \begin{equation}
 \label{pp2}
LR^{D}_{CA/SA} = LR^{q}_{CA/SA} = 1+ \frac{(1-q)}{q}e^{-2rt}.
 \end{equation}

We now consider the ratio  $\rho_{DS/D}(t)$ of two likelihood ratios.  One of them is the likelihood ratio for directional selection ($LR^{DS}_{CA/SA}$) when $q\neq s$; the other is the likelihood ratio for drift ($LR^{D}_{CA/SA}$), given by Eqn.  (\ref{pp2}). In both cases the two taxa are observed to have trait $X$.  
Thus,
 $$\rho_{DS/D}(t) = \frac{LR^{DS}_{CA/SA}}{LR^{D}_{CA/SA}}.$$
Notice that the numerator of $\rho_{DS/D}(t)$  involves a non-stationary process while the denominator characterizes a stationary process (with the stationary  probability of state $X$ being equal to $q$).  We assume $q \neq 0$ since otherwise the probability of observing trait $X$ at the present, according to the equilibrium (drift) model, is $0$ (under either CA or SA). 
 This framework allows us to derive the following proposition, which describes how the difference between directional selection and drift affects the degree to which a similarity favors common over separate ancestry:

 \begin{proposition}
 \label{rhodir}
 \mbox{ } 
 \begin{itemize}
 \item[(i)] For all $t>0$, $\rho_{DS/D}(t) >1$ if $1>q>s$ (selection against the trait) and $\rho_{DS/D}(t)<1$ if $q<s$ (selection for the trait).

\item[(ii)] If $q=s$ or $q=1$ then $\rho_{DS/D}(t)=1$ for all $t$.   
Moreover, $\lim_{t \rightarrow \infty} \rho_{DS/D}(t)$ and $\rho_{DS/D}(0)$ both equal $1$. 
\end{itemize}
 \end{proposition}

 	\begin{figure}[htb]
\centering
\includegraphics[scale=0.35]{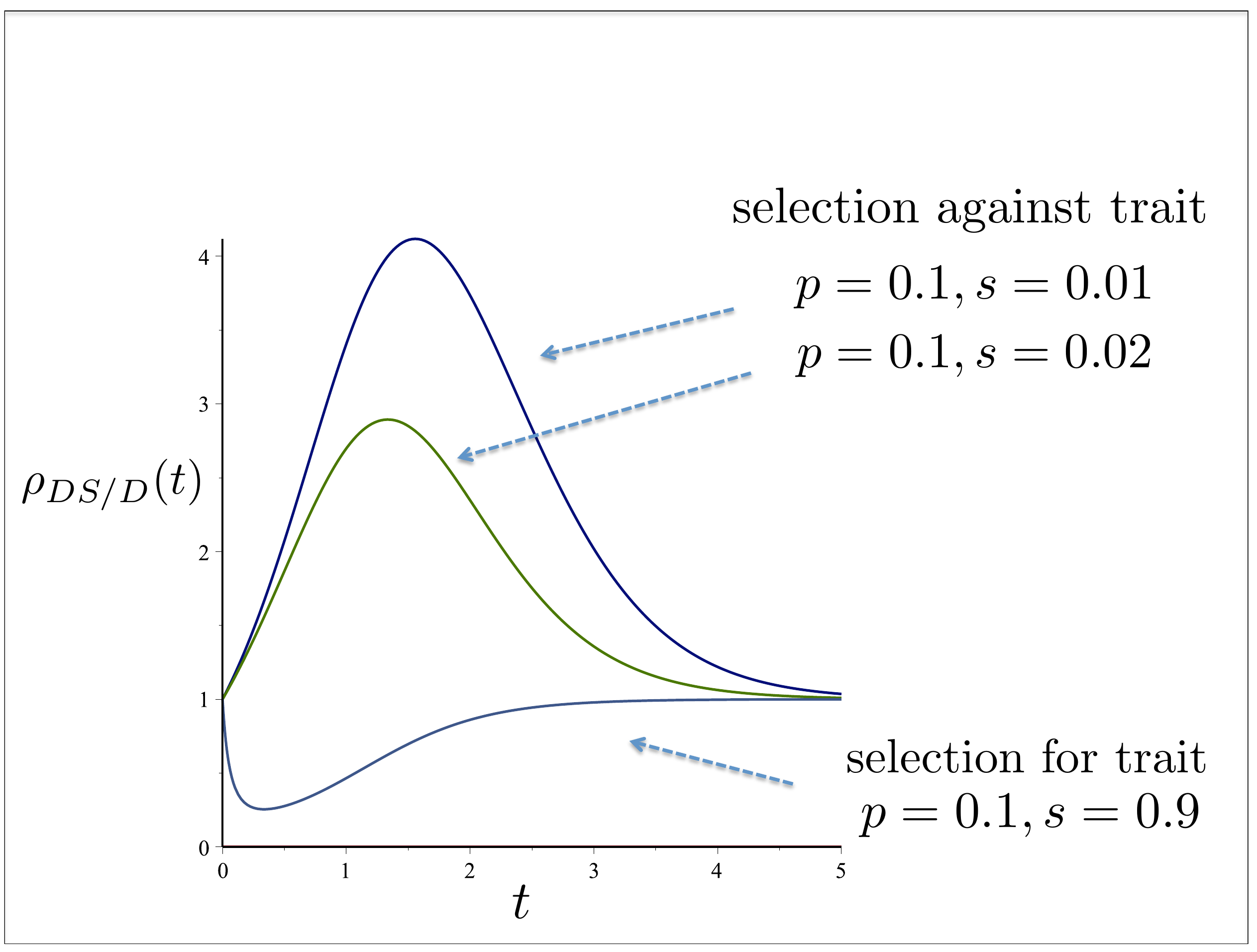}
\caption{
Three graphs of $\rho_{DS/D}(t)$, each of which compares directional selection and drift.  In each, an ancestor  ($t$ units in the past) has probability $q$ of being in state $X$.  In two of the curves, selection is compared with drift when there is selection against trait $X$, which is observed in the two leaves; in the third curve, selection is compared with drift when there is selection for the trait found in the leaves.  The time axis shows the expected number of state changes under the drift model.}
\label{figy}
\end{figure}  

 Figure~\ref{figy} shows three graphs of $\rho_{DS/D}(t)$ -- one compares selection with drift when there is selection for trait $X$ while the other two make the comparison when there is selection against trait $X$ (without loss of generality, we have taken $r=1$).  Proposition~\ref{rhodir} says that these behaviours are generic and provides a formal statement that accords with Darwin's idea that adaptive characters provide less support for CA than non-adaptive characters do. In fact, our result replaces Darwin's two types of similarity with three:  In an equal input model of directional selection, deleterious similarities are better than neutral similarities, and neutral similarities are better than adaptive similarities in terms of how much they favor CA over SA.

\section{Stablizing selection versus drift}
\label{stasec}

\begin{figure}[htb]
\centering
\includegraphics[scale=0.5]{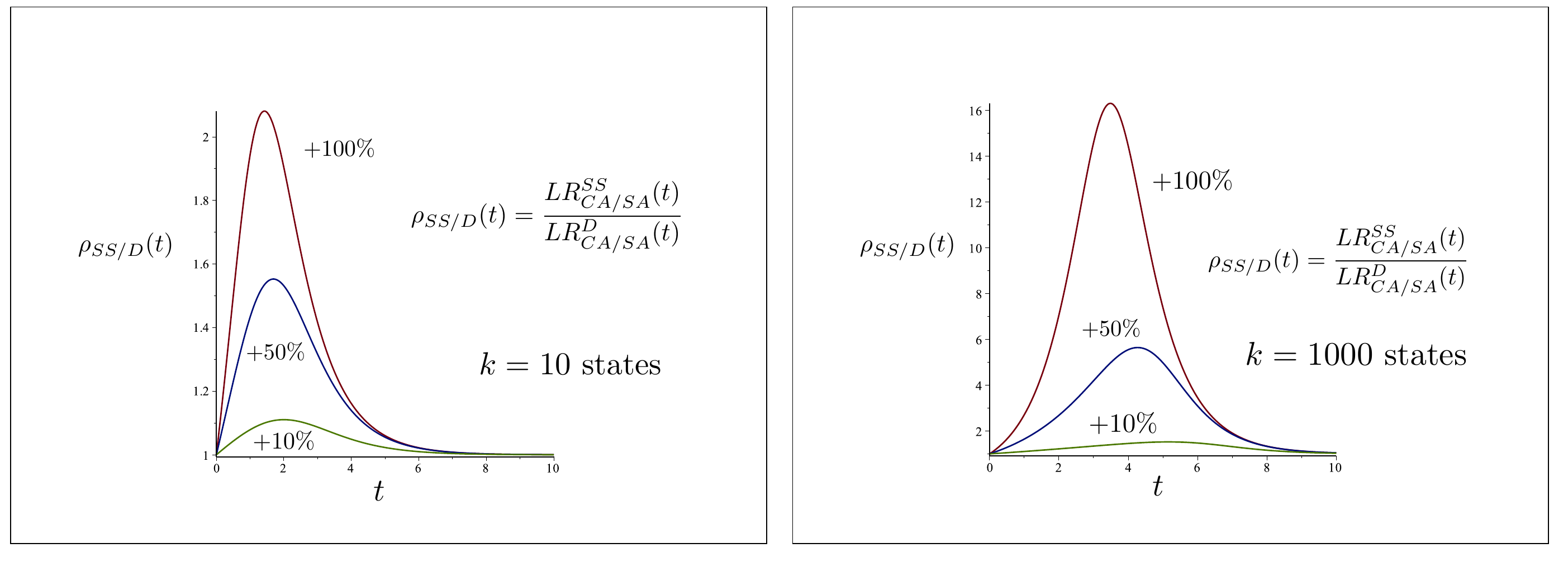}
\caption{The impact of stablising selection versus drift on the likelihood ratio of common ancestry versus separate ancestry for an equal input model with 10 states (left) and 1000 states (right), where all states are equally probable. The rate of leaving a state is 10\%, 50\% and 100\% higher for drift than for stabilising selection. The time axis shows the expected number of state changes under stablising selection. }
\label{figx}
\end{figure}

Suppose two taxa at the present share state $X$.
Consider the following ratio of likelihood-ratio values
 $$\rho_{SS/D}(t) = \frac{LR^{SS}_{CA/SA}}{LR^D_{CA/SA}}$$ 
 under the equal input model on $k$ states and two taxa.
 The numerator represents the likelihood ratio of common ancestry to separate ancestry when there is stabilizing selection (SS); the denominator represents that ratio when there is drift (D).
Suppose  $p$ is the stationary probability of state $X$, and that $r_D$ and $r_{SS}$ denote the substitution rates under drift and stabilising selection, where $r_D>r_{SS}$.\footnote{Here we depart from the usual conceptualization of stablizing selection in which a population has a bell-shaped distribution of some quantitative phenotype and the fitness of a trait value montonically increases as it gets closer to the population mean.}   From (\ref{sobeq}) it follows directly that:
$$\rho_{SS/D}(t)= \frac{1+ \left(\frac{1-p}{p}\right)e^{-2r_{SS}t}}{1+\left( \frac{1-p}{p}\right)e^{-2r_{D}t}}.$$
In this setting  stablizing selection inflates the likelihood ratio of CA over SA, and when the states have equal probability, the maximal inflation in the likelihood ratio grows according to a power law in the number of states:
\begin{proposition}
\label{rhostab}
\mbox{ }
\begin{itemize}
\item[(a)] For all $t>0$, $\rho_{SS/D}(t) >1$. Moreover, $\rho_{SS/D}(0) =1 = \lim_{t \rightarrow \infty} \rho_{SS/D}(t)$, and
$\rho_{SS/D}(t)$ has a unique critical point at some value $t^*>0$ where $\rho_{SS/D}(t)$ takes its global maximum value $M$.
\item[(b)] 
\begin{itemize}
\item[(i)] If  the substitution rate under drift is twice that for stabilising selection, then $M$ can be stated as an explicit closed-form function of $p$.
If, in addition, all $k$ states are equally probable, then 
$M =\frac{1}{2}(\sqrt{k}+ 1).$
\item[(ii)] More generally,  if the substitution rate under drift is $\tau>1$ times that for stabilising selection then we have the following asymptotic equivalence as $k$ becomes large:
$M \sim C_\tau \cdot k^{1-1/\tau}$, where the term  $C_\tau$ is independent of $k$ and is given by:
$C_\tau = \left (\frac{\tau-1}{\tau}\right)\cdot \left(\frac{1}{\tau-1}\right)^{1/\tau}.$
\end{itemize}
\end{itemize}
\end{proposition}
Notice in part (b)(ii) that as $\tau$ increases, the maximal value moves from being a small power of $k$ (e.g. square root when $\tau = 2$) towards
linear growth in $k$ (as $\tau \rightarrow \infty$). Notice also that when $\tau = 2$ then $C_\tau \cdot k^{1-1/\tau}  =\frac{1}{2}\sqrt{k}$
 in agreement with (b)(i).    Figure~\ref{figx} illustrates the behaviour of $\rho_{SS/D}(t)$ as a function of $k$ and the ratio $\tau = r_{D}/r_{SS}$.

It is useful to consider how this analysis of stabilizing selection versus drift is related to the $1/p$ argument described earlier.  Consider equation (\ref{condeq}), which holds regardless of how much time the common ancestry hypothesis says there is from taxa $A$ and $B$ back to their most recent common ancestor.  If drift and stabilizing selection assign the same value to $p$, how can the likelihood ratio of CA to SA be greater when there is stabilizing selection than when there is drift?  The answer is that $Pr_{CA}(A \mbox{ has } X |B \mbox{ has } X)$ has a higher value when there is stabilizing selection than when there is drift.  This is easily seen, since the models used for both processes  are stationary and  time-reversible. 

Proposition~\ref{rhostab} shows that Darwin overgeneralized when he said that adaptive similarities provide almost no evidence for common ancestry.  An adaptive similarity provides stronger evidence than a neutral similarity when the adaptive character evolves by stabilizing selection in an equal input model.  It is arguable that the ``framework of bones" that Darwin discussed and the near universality of the genetic code that we mentioned earlier each provide strong evidence for common ancestry because their evolution was governed by stabilizing selection. 
	
\section{Going beyond two taxa}
\label{beyond}

In Proposition~\ref{reich}, we described a very general sufficient condition for a trait shared by two taxa to favor common ancestry over separate ancestry.  Here we address a complication that arises when more than two taxa are considered.  The complication is that if more than two taxa have a common ancestor, there are different tree topologies that might connect those taxa to each other.  Thus, the hypothesis of common ancestry is a disjunction in which each disjunct is a different tree topology.  How is the likelihood of this disjunction to be compared with the likelihood of the separate ancestry hypothesis when $n$ leaf taxa all have the same trait value?  We address this question by identifying the tree topology that has the highest likelihood and the one that has the lowest likelihood; this means that the likelihood of the common ancestry disjunction must fall somewhere in between.

\begin{figure}[htb]
\centering
\includegraphics[scale=1.0]{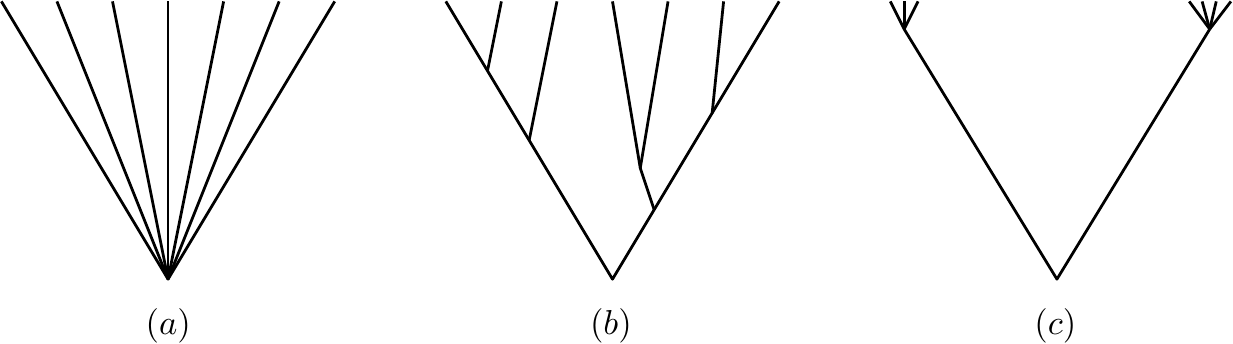}
\caption{(a) a star tree (b) a binary tree (c) a delayed tree}
\label{fig1ca}
\end{figure}

Suppose $n$ taxa share state $X$, and under CA  have a most recent common ancestor  $t$ time units in the past. 
Assume an equal input model of character state change in equilibrium (i.e. drift or stabilizing selection, but not directional selection) 
 in which state $X$ has stationary probability $p \neq 0$. We consider two extreme scenarios for the tree linking these $n$ taxa under CA:

\begin{itemize}
\item {\bf Star tree:}
This tree has all $n$ leaves adjacent to the root vertex, with edges of temporal length $t$.
For this tree it is readily verified that:
\begin{equation}
\label{starp}
LR_{\rm CA/SA} = p\left(1+\frac{1-p}{p} e^{-rt}\right)^n + (1-p)(1-e^{-rt})^n.
\end{equation}

\item {\bf Delayed tree:}
Consider a tree that has two edges both of length $t$, connecting the root to two leaves.   If each of the remaining $n-2$ leaves is attached to one or other of these leaves by edges of length zero, then we obtain a tree we call a `delayed tree'.   
For a delayed tree it is readily verified that:
$$LR_{\rm CA/SA}  =  \frac{p + (1-p)e^{-2r\cdot t}}{p^{n-1}}.$$
\end{itemize}
Notice that when $n=2$, simple algebra shows that the two expressions on the right for $LR_{\rm CA/SA}$ agree, and equal the expression in Eqn. (\ref{sobeq}), which is to be expected since for two leaves the star and delayed
tree are identical, and this is the only tree shape possible. Figure~\ref{fig1ca}  illustrates the star and delayed trees, on either side of a `typical' binary phylogenetic tree (note that  (c) shows only an approximation to a delayed tree, since edges of length zero are difficult to see!).

\begin{proposition}
\label{thm1}
\mbox{ } 
For any fixed parameters ($p, r, t$), where $p \neq 0, 1$, and $r, t>0$, the likelihood ratio $LR_{\rm CA/SA}$ is minimized when the underlying tree (under CA) is a star tree, and it is maximised by any delayed tree. 
\end{proposition}

This result has a number of immediate consequences:
\begin{enumerate}
\item
$LR_{\rm CA/SA}>1$ for all finite $t>0$. That is, a shared trait always favors CA over SA under the equal input model. 
\item 
When $n>2$ this proposition (and the formula  (\ref{starp}) for $LR$ for the star tree) improves on the lower bound from Sober and Steel (2014a, Proposition 1)  that stated the lower bound
$$LR_{\rm CA/SA}  \geq  1+\left(\frac{1}{p^{n-1}}-1\right)e^{-nr\cdot t},$$
which grows exponentially with $n$ when $p~< e^{-rt}$. However, the star bound is better since it grows exponentially with $n$
regardless of the size of $p\neq 0,1$.
\end{enumerate}

The result for the star tree in Proposition~\ref{thm1} might seem completely as expected.  Some caution is in order, however, since a related question led to the surprising finding that the star tree is the `extreme case' for certain types of equal input models, but not for others. More precisely, the star tree maximizes the mutual information between the states at the leaves and the root state for an equal input model on two equally probable states (Evans et al. 2000).  However, the star tree can fail to maximize mutual information when the equal input model has five or more equally probable states, provided  the number of leaves is sufficiently large, and the branch lengths lie in a certain range (Sly, 2011). 

\section{Conclusions}

	The idea that similarity is evidence for common ancestry has exceptions, but it holds in a very general circumstance, which we described by enumerating five assumptions.  Three of these are familiar from the literature on causal modeling:  intermediate probabilities, screening-off, and ancestor independence (Spirtes, Glymour, and Sheines 2000; Pearl 2009).   Two further assumptions are more specific to the literature on phylogenetic inference: cross-model homogeneity and cross-branch homogeneity.  We noted that the last of these is not an inevitable consequence of evolutionary theory.  If it is violated, a similarity can favor separate ancestry over common ancestry.  And even when the evolving trait obeys the five assumptions, tree topology complicates the likelihood comparison of common ancestry and separate ancestry; we   explained how an equal-input model permits the comparison to go forward when there are more than two leaf taxa.

Turning to the question of which similarities provide stronger evidence for common ancestry than which others, we began with a simple argument for the following thesis:  the sharing of trait $X$ among two leaf taxa provides stronger evidence for common ancestry the less probable it is that a taxon has trait $X$.  This is the $\nicefrac{1}{p}$ argument of Section~\ref{1psec}.  The main limitation of this argument is that it assumes that if two taxa have a common ancestor, their most recent common ancestor was in the very recent past.  Our analysis in Sections \ref{dirsec} and \ref{stasec} dropped that assumption, but our results show that the $\nicefrac{1}{p}$ argument was on the right track, at least in part.   When the selection process is directional selection, deleterious similarities are better than neutral similarities, and neutral similarities are better than adaptive similarities.  Darwin's comment that an adaptive similarity is ``almost valueless"  is correct if the adaptive similarity is due to directional selection.   However, when stabilizing selection is considered instead of directional selection, the situation is more subtle.  

\begin{figure}[htb]
\centering
\includegraphics[scale=0.75]{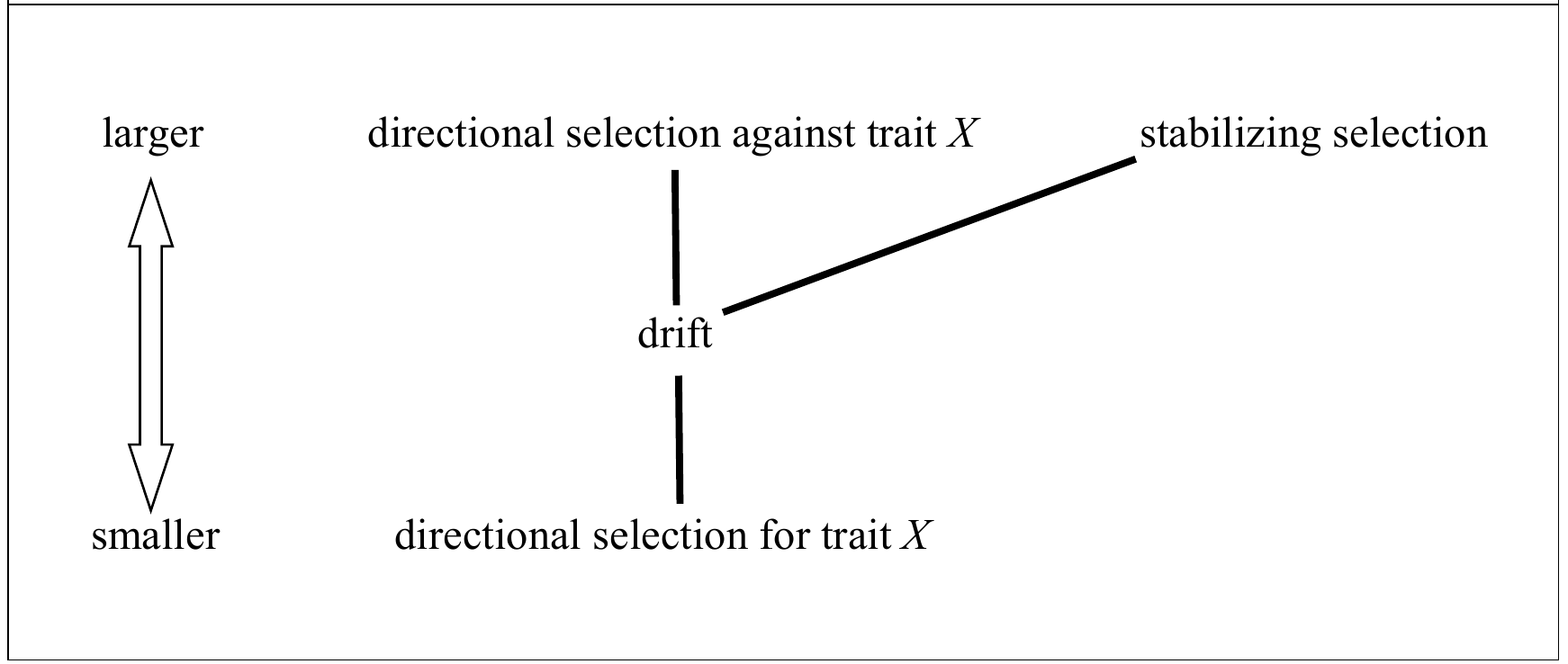}
\caption{A partial ordering of the likelihood ratios of Common Ancestry versus Separate Ancestry under four evolutionary processes.  In each case, two taxa are observed to share trait $X$.}
\label{fig_final}
\end{figure}

Under an equal-input model, adaptive similarities that are the product of stabilizing selection are better than neutral similarities. This means that the $\nicefrac{1}{p}$ argument is mistaken in this instance, since stabilizing selection and neutral evolution can assign the same probability to a taxon's having trait $X$.   Our results concerning the impact of different evolutionary processes on the ratio of the likelihoods of common ancestry and separate ancestry are summarized in  Figure~\ref{fig_final}.\footnote{ It is worth comparing this figure with Figure 3 in Sober and Steel (2014b), where the problem wasn't evidence for common ancestry, but the question of how much information the present state of a lineage provides about its ancestral state.  Sober and Steel use a Moran model framework to represent different evolutionary processes and take the present state of the lineage to be the frequency of an organismic trait.}

Although the $\nicefrac{1}{p}$ argument goes wrong in judging that stabilizing selection and drift are in the same boat when they assign the same probability to a taxon's having trait $X$, reducing the value of $p$ still plays a role in comparing these two processes.  The maximal extent to which stabilizing selection can favor CA over SA, compared with drift,  depends  on $p$;  Figure~\ref{figx} makes this plain for the special case of equally probable states.  In that case, $p = 1/k$ (where $k$ is the number of states) and the maximum degree to which stablizing selection  favors CA over SA, compared with drift, becomes large as $p$ becomes small -- this maximal ratio is described by a $\frac{1}{\sqrt{p}}$ relationship (when the  drift substitution rate is twice the stabilizing selection substitution rate) but moves closer to a $\nicefrac{1}{p}$ relationship as the ratio of these two substitution rates grows. 

Notice also that in both Figure~\ref{rhodir} and Figure~\ref{figx}, the likelihood advantage of one process over another sets in when $t>0$ and disappears as $t$ approaches infinity.
This is a pattern that should be expected in Markov processes that allow transitions from any state to any other state by some sequence of steps (Sober and Steel 2014b).

Although we used an equal-input model to represent stabilizing selection, the fact remains that stabilizing selection does not require an equal-input model.  This raises the question of how the difference between stabilizing selection and neutral evolution would affect the likelihood comparison of common and separate ancestry were stabilizing selection reconceptualized.   For example, consider an ordered set of character states $X_1, X_2, \ldots,  X_n$, where the probability of evolving from $X_i$ to $X_j$ depends on the value of $|i-j|$  -- the bigger the difference between $i$ and $j$, the smaller the value of $Pr(\mbox{descendant has } X_i| \mbox{ancestor has } X_j)$.
This ordering constraint reflects a type of  stabilizing selection.  The equal-input model provided a tidy solution to the problem we posed, but other models, like the one just described, need to be explored as well.  This caveat generalizes:  it is worth considering how the epistemological significance of different types of similarity is affected by varying model assumptions.  The present paper is not the end of the story.  









\section{Appendix}

{\em Proof of Proposition~\ref{reich}:}  The proof hinges on the following result (a general version of Reichenbach's theorem). 
\begin{lemma}
\label{lemre}
Suppose two events $E_1, E_2$ and a third variable $C$ take values in some discrete set $S$ of states. For any $x\in S$, let $C_x$ be the event that $C=x$. Suppose further that the following three conditions hold:
\begin{itemize}
\item[(i)] $E_1$ and $E_2$ are conditionally independent given $C_x$, for each $x \in S$;
\item[(ii)]$Pr(E_1|C_x)>Pr(E_1|C_y) \Rightarrow  Pr(E_2|C_x)>Pr(E_2|C_y)$ for all $x,y \in S$;
\item[(iii)]$Pr(E_1|C_x) \neq Pr(E_1|C_y)$ for some $x,y \in S$ with $Pr(C_x)>0$ and $Pr(C_y)>0$.
\end{itemize}   Then  $Pr(E_1 \& E_2) > Pr(E_1)Pr(E_2)$.
\end{lemma}
{\em Proof:}  By a standard trick, a short proof is possible thanks to the following convenient equation (which follows from assumption (i) and algebra):
\begin{equation}
\label{sumsum}
\sum_{x\in S, y\in S}\Delta_1(x,y)\Delta_2(x,y)Pr(C_x) Pr(C_y) =  2[Pr(E_1 \&E_2) - Pr(E_1)Pr(E_2)]
\end{equation}
where
 $\Delta_i(x,y)= Pr(E_i|C_x) - Pr(E_i|C_y)$,
coupled with the observation that each summand in (\ref{sumsum}) is non-negative (by (ii)), and therefore the sum is strictly positive (by (iii)).
\hfill$\Box$

With this result in hand, the proof  of Proposition~\ref{reich} now follows, by taking
$E_1$ to be the event that taxon $A$ is in state $X_k$, $B$ to be the event that taxon $B$ is in state $X_k$, and $C$ to be the
state of the most recent common ancestor of $A$ and $B$ under CA.  
Lemma~\ref{lemre}, together with Assumptions 2(i), 3, and 5,  
shows that:
\begin{eqnarray}
Pr(\mbox{taxa $A$ and $B$ are in state $X_k | CA$}) > & \\
Pr(\mbox{taxon $A$ is in state $X_k| CA$})Pr(\mbox{taxon $B$ is in state $X_k | CA$}),
\label{ineqx}
\end{eqnarray}
and Assumptions 2(ii), 2(iii), and 4 imply that  $Pr(\mbox{taxa $A$ and $B$ are in state $X_k | SA$})$ is equal to $Pr(\mbox{taxon $A$ is in state $X_k | SA$})Pr(\mbox{taxon $B$ is in state $X_k| SA$})$, so that the latter term,  by Assumption 1,  is equal to (\ref{ineqx}).

\bigskip

{\em Proof of Proposition~\ref{rhodir}:}  Notice that we can write:
 $$\rho_{DS/D}(t) = \frac{q(s+(1-s)\theta)^2 + (1-q)s^2(1-\theta)^2}{(s+(q-s)\theta)^2(1+(\frac{1}{q} -1) \theta^2)},$$
 where $\theta = e^{-rt}$.
 Let $\Delta$ denote the numerator of $\rho_{DS/D}(t)$ minus the denominator. 
 Then tedious but straightforward algebra shows that:
 $$\Delta=\left(\frac{1}{q}-1\right) \theta (q-s)[s(1-2\theta^2 + \theta^3) + q(1-\theta^3)].$$
 Now, since $\theta \in (0, 1)$ for $t>0$, we have that $(1-2\theta^2 + \theta^3)>0$, and $(1-\theta^3)>0$ so the term in the square brackets in the last equation is strictly positive (since $q>0$) and $\frac{1}{q}-1>0$ (unless $q=1$ in which case $\rho_{DS/D}(t)=1$ for all $t$).
 
Consequently, the sign of $\Delta$ when $q\neq s$ and $q\neq 1$ is exactly the sign of $(q-s)$, which gives the result claimed (since $\rho_{DS/D}(t)$ is greater or smaller than 1 precisely when $\Delta$ is positive or negative).  The proof of part (ii) of Proposition~\ref{rhodir} is straightforward.

\hfill$\Box$

\bigskip

{\em Proof of Proposition~\ref{rhostab}:} For part (a), consider the difference $\delta$ between the numerator and denominator of $\rho_{SS/D}(t)$. Then,
$$\delta = \left(\frac{1-p}{p}\right)(e^{-2r_{SS}t}- e^{-2r_{D}t})>0,$$
for $t>0$ since $r_D>r_{SS}$. Thus, $\rho_{SS/D}(t)>1$ for all $t>0$.

Now, any solution to the  equation $\frac{d}{dt}\rho_{SS/D}(t)=0$, satisfies:
\begin{equation}
\label{unique}
re^{st}-se^{rt} + (r-s)\left(\frac{1-p}{p}\right) = 0,
\end{equation}
where, for brevity  we write  $r=r_D$ and $s=r_{SS}$ (so $r>s$) here and in what follows.
To see that Eqn.~(\ref{unique}) has a unique solution, notice that the left-hand-side is strictly positive when $t=0$, and tends to $-\infty$ as $t$ grows; and since the derivative
of the left-hand-side with respect to $t$ is $rs(e^{st}-e^{rt})$ which is strictly negative for all $t>0$, the left-hand-side cuts the $t-$axis exactly once, and so equals zero for a unique value $t^*$, as claimed.

When $r=2s$, Eqn.~(\ref{unique})  becomes (upon division through by $s$) the following quadratic equation for $x=e^{st}$:
$2x - x^2 = -\left(\frac{1-p}{p}\right),$
which has a unique solution for $x>1$, namely,
\begin{equation}
\label{eqx}
x = 1+ \sqrt{1+ (1-p)/p}
\end{equation}
and from this we obtain an explicit expression for $M$, namely:
\begin{equation}
\label{pxpeq}
M= \frac{1+(1-p)/px}{1+(1-p)/px^2},
\end{equation}
 where $x$ is given by Eqn. (\ref{eqx}).
In case $p = 1/k$, Eqn.~(\ref{eqx}) gives:
$$x = \sqrt{k}+1,$$ from which Eqn. (\ref{pxpeq}) becomes, upon simplification:
$$M =\frac{1}{2}(\sqrt{k}+1).$$

For part (ii), let  $y=e^{-st}$.  The assumptions that $r= \tau \cdot s$ and $p=1/k$ imply that 
$\rho_{SS/D}(t) =\frac{1+(k-1)y}{1+(k-1)y^\tau}$. This expression is maximized  at the $t$ value for which $y$ satisfies the equation:
$$(k-1)(\tau-1) y^\tau + \tau y^{\tau-1} - 1=0.$$
The solution to this last equation in the range $(0,1)$ is (asymptotically as $k$ grows) given by
$y \sim \left[\frac{1}{(k-1)(\tau-1)}\right]^{1/\tau}$, from which part (ii) now follows.
\hfill$\Box$

\bigskip

{\em Proof of Proposition~\ref{thm1}:}   Suppose that $T$ is a rooted tree on $n\geq 2$ leaves, where the root vertex $\rho$ is the recent common ancestor of the leaves. 
Suppose that $T$ is not a star tree.  Then $n\geq 3$, and  $T$ has a vertex $v$ that is adjacent to the root of $T$, and which has edges to at least two other pendant subtrees $T_1, \ldots, T_k$. Let $l_1$ and $l_2$ denote the lengths of the edges that connect the root to $v$ and $v$ to the root of $T_1$, respectively. 
Consider the tree $T'$ obtained by reattaching $T_1$ directly to the root of $T$, by an edge of length $l_1+l_2$.  We will show that $T'$ has a lower probability that all its leaves are in state $X$ than $T$ does.  It then follows that only the star tree minimizes this probability.

Let $E_1$, $E_2$ and $F$ denote, respectively, the events that all the leaves in $T_1$, in $T_2$--$T_k$, and in the remainder of $T$, are in state $X$, and let $E$ denote the conjunction of these three events (i.e. the event that all the leaves of $T$ are in state $X$). 
Let $E'_1$, $E'_2$ and $F'$ and $E'$ denote, the corresponding events for tree $T'$.
If $Y_\rho$ denotes the state at the root of each tree, then, by the law of total probability:
\begin{equation}
\label{fue1}
Pr(E) = \sum_{y} Pr_y(E_1 \&E_2\&F)Pr(Y_\rho=y),
\end{equation}
and
\begin{equation}
\label{fue2}
Pr'(E) = \sum_{y} Pr'_y(E'_1 \&E'_2\&F')Pr'(Y_\rho=y),
\end{equation}
where $Pr$ and $Pr'$ denote  probabilities computed on $T$ and $T'$ respectively, and
where $Pr_y$ and $Pr'_y$ denote (conditional) probabilities  computed on $T$ and $T'$ respectively, conditional on the root-state event $Y_\rho=y$.
 Notice that $Pr(Y_\rho=y) = Pr'(Y_\rho=y)$.
Also,  $Pr_y(E_1 \&E_2 \&F)=Pr_y(E_1 \&E_2)Pr_y(F)$  and  $Pr'_y(E'_1 \&E'_2 \&F)=Pr'_y(E'_1 \&E'_2)Pr'_y(F')$ since $Y_\rho$ screens-off 
$E_1\&E_2$ from $F$ in $T$,  and also $E'_1\&E'_2$ from $F'$ in $T'$.  Moreover, 
$Pr_y(F) = Pr_y'(F')$ for all $y$. Thus, to establish that  $Pr'(E')<Pr(E)$, it suffices, by (\ref{fue1}) and (\ref{fue2}),  to  show that, for every state $y$: 
\begin{equation}
\label{eqxy}
Pr_y(E_1 \&E_2)> Pr'_y(E'_1 \&E'_2).
\end{equation}
Notice that $E_1$ and $E_2$ become conditionally independent once we specify the state at vertex $v$, which we denote as $Y_v$ (this variable also screens-off $Y_\rho$ from these $E_1$ and $E_2$). 
Now, for 
 $i=1$ and $i=2$, we have
$Pr_y(E_i|Y_v = X)> Pr_y(E_i|Y_v = X')$ 
for any state $X' \neq X$; moreover, the nature of the equal input model ensures that
$Pr_y(E_i|Y_v = X')= Pr_y(E_i|Y_v = X'')$  for any two states $X', X''$ that are different from $X$.
In addition, $Pr_y(Y_v=X')>0$ for $X' = X$ and for at least one other state $X' \neq X$ (since $p \neq 0,1$).  
Thus, we have satisfied conditions (i)--(iii) in the general version of Reichenbach's theorem (Lemma~\ref{lemre}, taking $C=Y_v$) to deduce that:
\begin{equation}
\label{eqxy2}
Pr_y(E_1 \& E_2)> Pr_y(E_1)Pr_y(E_2).
\end{equation}
Turning to  $Pr'_y(E'_1 \&E'_2)$ we have:
$$Pr'_y(E'_1 \&E'_2) = Pr_y'(E'_1) Pr'_y(E'_2).$$
Now, considering the right-hand-side of this last equation, notice that:
$$Pr_y'(E'_1)  = Pr_y(E_1) \mbox{ and } Pr_y'(E'_2) =Pr_y(E_2). $$
Thus, $Pr'_y(E'_1 \&E'_2) = Pr_y(E_1)Pr_y(E_2)$ 
which, by (\ref{eqxy2}), establishes the required Inequality (\ref{eqxy}).

\bigskip

For the result concerning the delayed tree we use  an equivalent description of the equal input  model sometimes referred to as the `Fortuin-Kasteleyn'  random cluster model (see Section 2.1 of Matsen, Mossel and Steel 2008).  Let $C =1,2, \ldots, n$ be the number of clusters (blocks of the partition of the set of leaves of $T$ induced by an independent Poisson process that acts with intensity $r$ along the edges of the tree; the partition regards two leaves as being in the same block if the path between them does not cross an edge on which the Poisson event has occurred). Here $r$ is the  substitution rate, divided by 1 minus the sum of the squares of the stationary probabilities of the states.
Then if $\psi_X$ denote the probability that, under the equal input model, all $n$ leaves of $T$ are all in state $X$, and  if $p$ denotes the stationary probability of state $X$, the random cluster description allows us to write $\psi_X$ as follows:
\begin{equation}
\label{psieq}
\psi_X = \EE[p^C] = \sum_{i =1}^n Pr(C=i) p^i.
\end{equation}
Thus, 
$$\psi_X  \leq p \cdot  \PP(C=1) + p^2 \cdot \PP(C>1).$$
Noting again that $\PP(C=1) = e^{-rL},$ we get:
\begin{equation}
\label{gel2} 
\psi_X \leq pe^{-rL} + p^2(1-e^{-rL}) = p^2 + p(1-p)e^{-rL}.
\end{equation}
Now, $L \geq 2t$ with equality if and only if the tree is a delayed tree.   It follows that
$\psi_X \leq p^2 + p(1-p)e^{-2r\cdot t}$.  Dividing this by again by $p^n$ we arrive at
the upper bound on $LR$ given by the expression for the delayed tree.
\hfill$\Box$

\section{References}

\hangindent=\parindent
\hangafter=1
\noindent
Crick, F. [1957]: `The Origin of the Genetic Code', {\em Journal of Molecular Biology} {\bf 38}, pp. 367--379. 

\hangindent=\parindent
\hangafter=1
\noindent
Darwin, C. [1859]: {\em  On the Origin of Species by Means of Natural Selection}, London: Murray.

\hangindent=\parindent
\hangafter=1
\noindent
Evans, W., Kenyon, C., Peres, Y. and Schulman, L.J.  [2000]: `Broadcasting on
trees and the Ising model', {\em Advances in Applied Probability}, {\bf 10}, pp. 410--33.

\hangindent=\parindent
\hangafter=1
\noindent
 Hacking, I. [1965]: {\em  Logic of Statistical Inference}, Cambridge: Cambridge University Press.

\hangindent=\parindent
\hangafter=1
\noindent
Knight, R., Freeland, S., and Landweber, L. [2001]: `Rewiring the keyboard -- Evolvability of the genetic code', {\em Nature Reviews - Genetics}  {\bf 2}, pp. 49--58.

\hangindent=\parindent
\hangafter=1
\noindent
  Lemey, P., Salemi, M. and Vandamme, A-M. [2009]: {\em  The phylogenetic handbook: A practical approach to phylogenetic analysis and hypothesis testing. 2nd ed.}, Cambridge: Cambridge University Press.

\hangindent=\parindent
\hangafter=1
\noindent
Lewens, T. [2015]: `Backwards in retrospect.' {\em  Philosophical Studies} (in press). 

\hangindent=\parindent
\hangafter=1
\noindent
Matsen, F. A., Mossel, E. and Steel, M. [2008]: `Mixed-up trees: The structure of phylogenetic mixtures', {\em  Bulletin of Mathematical Biology}, {\bf 70(4)}, pp. 1115--1139.

\hangindent=\parindent
\hangafter=1
\noindent
 Pearl, J. [2009]: {\em Causality: Models, Reasoning, and Inference}, 2nd edition, 2009. New York: Cambridge University Press.

\hangindent=\parindent
\hangafter=1
\noindent
Reichenbach, H.  [1956]: {\em The Direction of Time.}  Berkeley: University of California Press.

\hangindent=\parindent
\hangafter=1
\noindent
Semple, C. and Steel, M. [2003]: {\em  Phylogenetics.} Oxford: Oxford University Press.

\hangindent=\parindent
\hangafter=1
\noindent
Sly, A. [2011]: `Reconstruction for the Potts model',  {\em Annals of Probability}, {\bf 39}, pp. 1365--1406.

\hangindent=\parindent
\hangafter=1
\noindent
Sober, E. [2008]:  {\em Evidence and Evolution -- the Logic Behind the Science.}  Cambridge: Cambridge university Press.

\hangindent=\parindent
\hangafter=1
\noindent
Sober, E. [2011]: {\em Did Darwin write the Origin Backwards?},  Amherst, New York: Prometheus Books.

\hangindent=\parindent
\hangafter=1
\noindent
Sober, E. [2015]: {\em  Ockham's Razors -- A User's Manual}, Cambridge: Cambridge University Press.

\hangindent=\parindent
\hangafter=1
\noindent
 Sober, E. and Steel, M. [2014a]:  `Is natural selection evidence for common ancestry?' (manuscript).

\hangindent=\parindent
\hangafter=1
\noindent
Sober, E. and Steel, M. [2014b]: `Time and knowability in evolutionary processes',
{\em  Philosophy of Science}, {\bf 81}, pp. 537--557.

\hangindent=\parindent
\hangafter=1
\noindent
Spirtes, P.,  Glymour, C. and Sheines, R.  [2000]:  {\em Causation, Prediction, and Search}, 2nd edition, Cambridge, MA: MIT Press.

\end{document}